# ORSAY LECTURES ON CONFINEMENT (I)

by

**Vladimir N. Gribov**\*

L. D. Landau Institute for Theoretical Physics

Acad. of Sciences of the USSR, Leninsky pr. 53, 117 924 Moscow, Russia

and

KFKI Research Institute for Particle and Nuclear Physics

of the Hungarian Academy of Sciences

H – 1525 Budapest 114, P.O.B. 49, Hungary

and

Laboratoire de Physique Théorique et Hautes Energies\*\*

Université de Paris XI, bâtiment 211, 91405 Orsay Cedex, France

LPTHE Orsay 92/60

June 1993

---

\* Supported in part by Landau Institute-ENS Département de Physique exchange program

\*\* Laboratoire associé au Centre National de la Recherche Scientifique

# 1. The theory of supercharged nucleus

This talk is just introductory. Literally it is not about confinement, but it is important since it is closely connected to the theory of quark confinement. In fact, this theory deals with two different problems. The first one is, how to confine particles, i.e. the problem of producing forces strong enough to prevent quarks from separating, which will be discussed later. But there is also another, very severe problem.

We know, that quarks are light, almost massless particles. The question is, how to bind massless particles in a volume which is much smaller than the Compton wave length. Usually this is not easy to do, because the wave function is decreasing exponentially like $e^{-\sqrt{m^2-\omega^2}r}$, which in the best case can be $e^{-mr}$. But if the mass is very small, the state is very broad, and it is in complete contradiction with what we know about hadrons.

The mechanism of supercharged nuclei, which we will discuss in the present talk, seems to be a unique possibility to bind a particle in a small region in space. The theory of the supercharged nucleus is very old. It was initiated in the forties by the work of Pomeranchuk and Smorodinsky [1], and it became very well developed, with no unsolved questions left. Since we, however, want to apply this theory to quarks, we will talk about it in a way slightly different from what people are used to, combining the picture of the Dirac sea in external field and the language of Feynman's Green function.

The problem of the supercharged nucleus is the following. If there is a nucleus $N_Z$ with a charge $Z$, and if $Z$ is larger than a critical value $Z_c$

$$Z > Z_c$$

(theoretically $Z_c = 137$, practically it is around $Z_c \sim 180$), then this nucleus will decay to an atom with a charge $Z - 1$ and a positron:

$$N_Z \to A_{Z-1} + e^+ \ .$$

This atom can be stable or unstable. If it is unstable, it can decay again - up to a situation, when the total charge of the atomic state $A_{Z-n}$ becomes sufficiently small. This is a peculiar thing : it means, that the nucleus behaves like a resonance. It does not exist in the nature freely, but it exists inside the atom. In this sense it is analogous to what we know about the quarks. This, of course, is not confinement, but in some respects it is not so different.

Indeed, the nucleus has a baryon number B and a lepton number zero. But this state with lepton number 0 does not exist, there are only states for which the lepton number equals unity. In this sense, it is a confinement of states with zero lepton number.

Theoretically the described problem is very well defined. If we want to understand the new type of atomic states, we have to consider the interaction of the electron with the nucleus, taking into account all possible interactions.

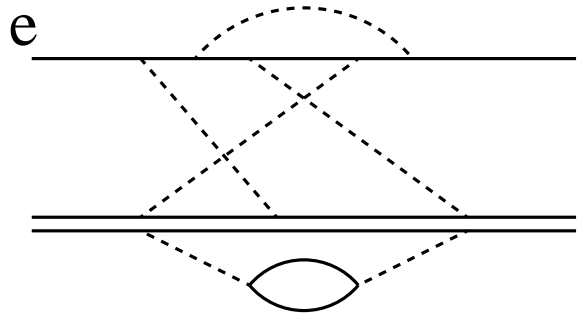

In this case there are some simplifications. First,

$$\alpha \ll 1 \;, \tag{1}$$

but

$$\alpha Z \sim 1 \;. \tag{2}$$

The third important simplification is that the ratio of the electron mass $m_e$ and the nucleon mass $m_N$ is much smaller than unity :

$$\frac{m_e}{m_N} \ll 1 \;. \tag{3}$$

Because of these conditions, the problem can be solved exactly. First, due to (1), we can neglect all the corrections to electron propagation, since they would be of the order of $\alpha$, and leave those diagrams which contain lines connecting electron and nucleus lines and radiative correction to the nucleus line. But also, because of (3), the recoil of the nucleus due to photon emission will be very small, and instead of writing this interaction and taking into account all the sequences, including recoil, we can write that this is equal to the electronic Green function in the external field of the nucleus, multiplied by the nuclear Green function with radiative corrections

$$\cdots \longrightarrow \cdots = G_e(A)\, G_N \tag{4}$$

Here $G_e(A)$ is the Green function of the electron in the external field ; the Green function of the nucleus $G_N$, which corresponds to the diagram a contains a lot of radiative corrections. We have to include loops like b since they are of the order of $\alpha Z$, and diagrams c being $\alpha Z^2$. Fortunately, all this turns out to be very simple. The reason is, that all the photons in these diagrams cause no recoil for the nucleus. It means, that we can write the following :

$$G_N = G_N^0 e^{\bigcirc + \boxtimes} \tag{5}$$

where everything can be factorized. $\bigcirc$ will be the symbolic expression for the electrostatic energy of the nucleus, and $\boxtimes$ gives us the mass renormalization for the nucleus due to electron vacuum polarization. And thus we have an exact description for $G_N$. Further, the first step is, of course, to calculate the electron Green function in an external field, which also defines the nucleus mass renormalization entering the exponent in (5).

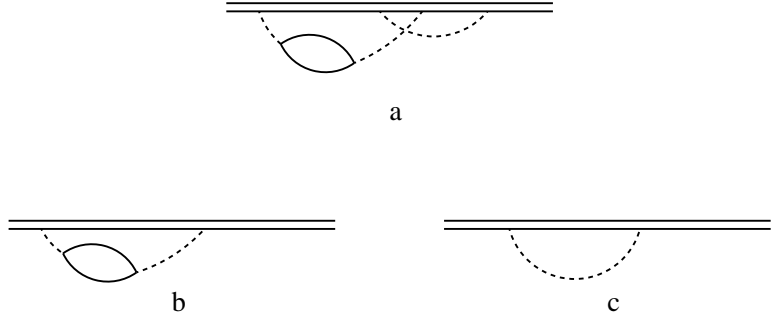

This means, that with the knowledge of the Green function of the electron in the Coulomb field, we can calculate everything.

The equation for the electron Green function is

$$\widehat{\nabla} G_e = -i\delta \tag{6}$$

$$\gamma_\mu \left(\partial_\mu - iA_\mu\right) G_e = -i\delta \tag{7}$$

Since the potential does not depend on the time, we can always write

$$G_e(x_2, x_1) = \int \frac{d\omega}{2\pi i} e^{-i\omega(t_2 - t_1)} \psi_\omega(r_1) \bar{\psi}_\omega(r_2) \tag{8}$$

and, because of that, we have

$$\gamma_\mu \left(\partial_\mu - iA_\mu\right) \psi_\omega = 0 \ , \tag{9}$$

or, in a more convenient way,

$$[\alpha_j p_j + m\gamma_0 - (\omega - A)] \psi_\omega = 0 \ . \tag{10}$$

Before discussing the solution of the equation, let us consider the well-known features of the spectrum of this system.

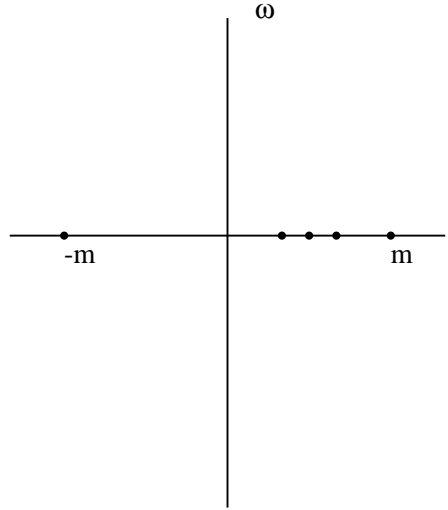

There are cuts going from $m$ to $\infty$ and from $-m$ to $-\infty$, corresponding to the continuous spectrum, and sequences of poles corresponding to Coulomb levels.

Now - how does the critical phenomenon come in ? If we increase $Z$, the pole which corresponds to the atom in the ground state moves to the left, passing zero without any troubles (for a finite-size nucleus) and reaches the point $-m$ at the value $Z = Z_c$. With the further growth of $Z > Z_c$ the position of the level $\omega_0$ is going to the complex plane and the state becomes unstable. However, at this point we come to a paradox : this $\omega_0$ is supposed to be the energy of an atomic state, i.e. the energy of the atomic state becomes complex, contradicting the physics we have expected. Indeed, we thought that the nucleus was unstable and would decay on a stable atom and a positron.

This means, that the described simple solution has to be essentially changed, since, apparently, the Green function $G_e(A)$ of the electron does not reflect the whole physics, and the features of $G_N$ in (4) turn out to be important. We will show, in fact, that there is a cancellation between $G_e(A)$ and $G_N$.

Let us consider now, what is happening from the point of view of the Dirac sea. From $m$ to $\infty$ and from $-m$ to $-\infty$ there is a continuous spectrum of electron states in the Coulomb field. Also, as we said before, there is a resonance state with complex energy $\omega_0 + i\Gamma/2$, where $\omega_0 < -m$. What is the physics of this resonance ?

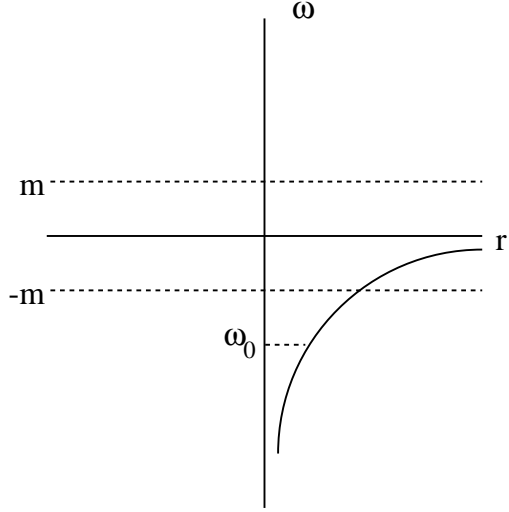

The Dirac equation can be written in the form

$$(T + A)\psi = 0 \tag{11}$$

where T is the kinetic energy and $A$ the potential energy. This kinetic energy can have two signs $T = \pm\sqrt{p^2 + m^2}$, and therefore the total energy can be quasi-classically either

$$\sqrt{p^2 + m^2} + A(r) = E \tag{12}$$

or

$$-\sqrt{p^2 + m^2} + A(r) = E \; . \tag{13}$$

Obviously, (12) can be fulfilled only close to the origin. In this case the potential is negative, the square root is positive, and the total energy is negative. Classically, the electron will be stopped at a return point $r_1$, where the energy $E$ is the sum of the potential and the mass $m$ :

$$m + A(r_1) = E \tag{14}$$

This means, that in this region there is a normal wave function which oscillates at $r < r_1$. After that it has to decrease exponentially because of the absence of classical trajectories. For the equation (13) the situation is the opposite, it can exist only at large distances, since the potential at large distances is small, and the energy is negative ; consequently, there will be another return point $r_2$, where

$$-m + A(r_2) = E \qquad (15)$$

Again, there will be a plane wave, i.e. normal oscillation at $r > r_2$ with a subsequent exponential decrease due to the absence of classical motion between the points $r_1$ and $r_2$

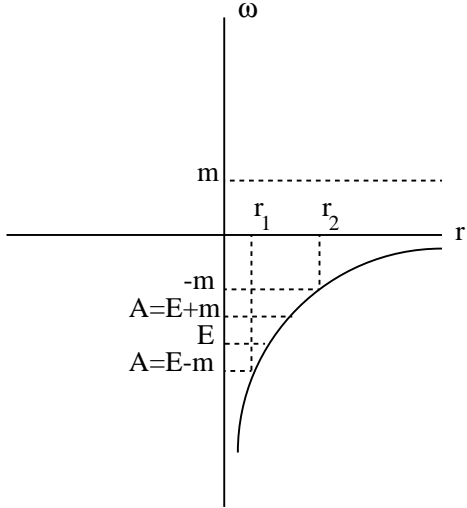

The situation is somewhat similar to that with a barrier in the non-relativistic problem :

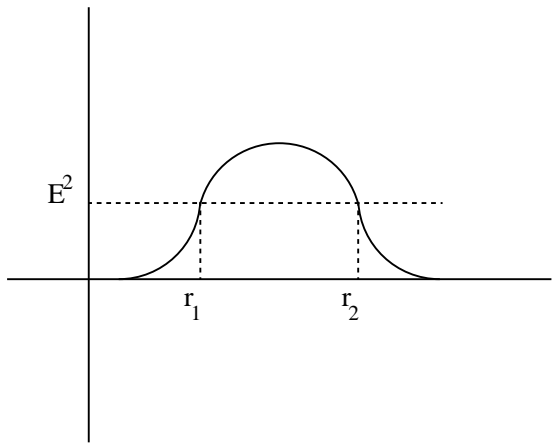

This barrier-type behaviour can be easily seen, if one writes a second order equation instead of the first order one. Indeed, let us re-write the Dirac equation in the form

$$\left(\nabla^2 + \frac{1}{2}F_{\mu\nu}\sigma_{\mu\nu}\right)\psi = 0 \ . \tag{16}$$

Here

$$\nabla^2 - \left(\partial_t^2 - iA\right)^2 = -\frac{1}{r}\partial_r^2 r - \omega^2 + 2\omega A - A^2 \ . \tag{17}$$

In this equation the effective potential energy is $2\omega A - A^2$. If $A$ is negative, the first term $2\omega A$ corresponds to attraction at large distances, and the second term $-A^2$ to repulsion at short distances.

It is natural to expect, that this negative energy state is just a resonance. But : in the previous discussion we forgot about the Pauli principle. We have been talking only about one state and not about the Dirac sea. However, Dirac sea means, that all these normal negative levels have to be occupied.

In this case, if our particle will try to go out and pass through this barrier, there will be no place for it. When we find, that the Green function $G_e$ has a complex singularity, this reflects the fact, that we did not say up to now, what type of Green function we are using.

Imagine, that the Dirac sea is not filled up. In this case it would be natural for our state to have complex energy. On the other hand, we are used to the fact, that the Feynman Green function reflects the Pauli principle in the correct way, and therefore, if we are looking for the Feynman Green function, we expect to find the proper answer.

Suppose now, that we will calculate the Feynman Green function. Still, we said that there is a general solution for the Green function with a complex pole. The question is, how this happens.

The answer is essentially very simple. Indeed, we are looking for a Green function which is a sum of diagrams of the type

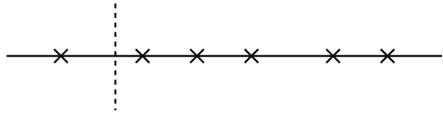

$$(18)$$

If the Feynman Green function contained only positive energy propagating in positive time, all the energy denominators in (18)
$$\frac{1}{\varepsilon - \sqrt{m^2 - p^2}}$$
would be real at negative total energy. But the Feynman Green function contains also negative energy, propagating in negative time. It corresponds to the so-called $Z$-diagram

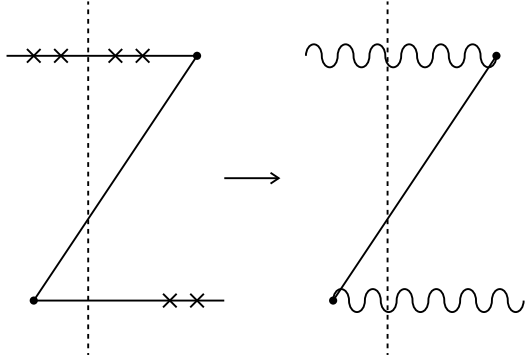

Perturbatively this diagram contains also only positive energy intermediate states. However, if the incoming and outgoing lines will correspond to negative energy bound states, then the three-particle intermediate state containing two negative energy particles and one positive energy particle can have negative total energy :

$$\frac{1}{\varepsilon - \sqrt{m^2 + p^2} - 2\varepsilon} = -\frac{1}{\varepsilon + \sqrt{m^2 + p^2}} \ .$$

This means, that the Feynman Green function in this case obtains an imaginary position of the pole in contradiction with the Pauli principle. (This contradiction is obvious from the second of the Z-diagrams, since we have there two particles in the same state at the same time). It is also clear, that this diagram reflects not the decay of a state, but that of the vacuum.

Let us see, how the Green function of the nucleus will recover the Pauli principle. Consider the diagram

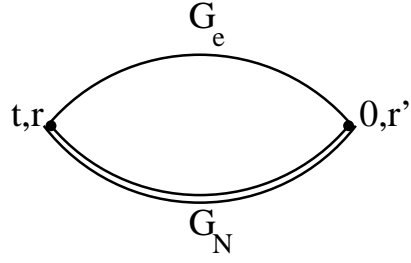

The Green function of the electron is in this case

$$G_e(t, r, r') = \int \frac{d\omega}{2\pi i} e^{-i\omega t} \psi_\omega(r) \bar{\psi}_\omega(r') \quad . \tag{19}$$

In order to calculate it, we have to write the following contour of integration (for $Z < Z_c$)

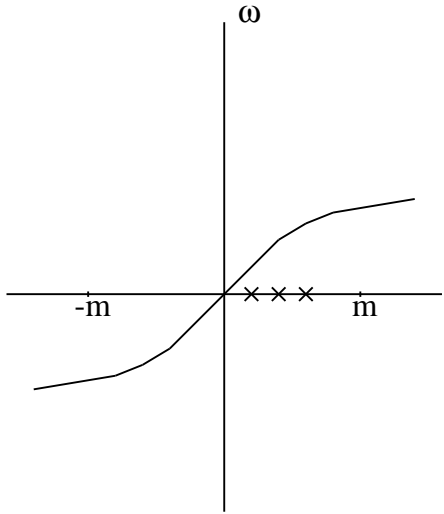

With $Z$ growing, the pole will move (as indicated by the dotted line) to the complex plane,

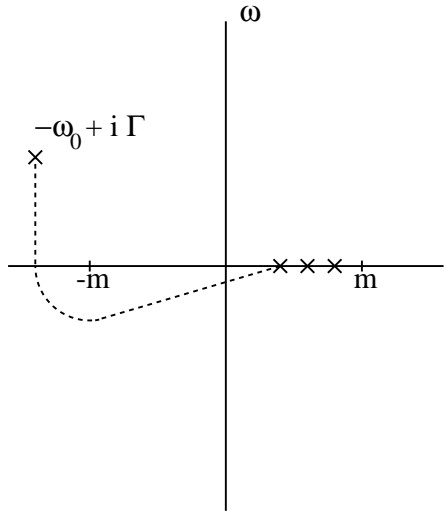

and we will have to change the contour of integration:

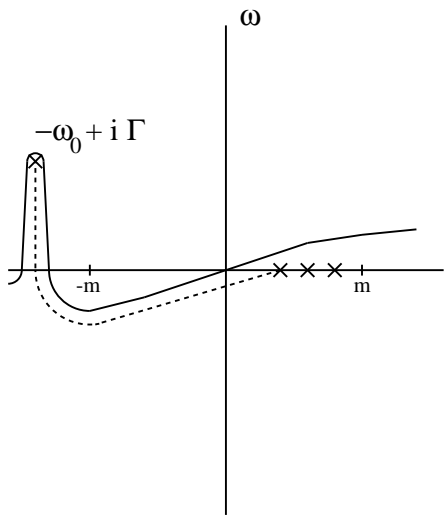

As we see, the Green function will contain a growing exponent which will be defined by the imaginary part of the pole. At large $t$ we will be left with an unstable pole contribution

$$G_e(t,r,r') = \int \frac{d\omega}{2\pi i}\ e^{-i\omega t}\psi_\omega(r)\ \bar{\psi}_\omega(r') \sim e^{\Gamma t - i\omega_0 t}\ .$$

Obviously, the total of the diagrams can not contradict the Pauli principle, and due to this fact there has to be a cancellation between the electron Green function and the Green function of the nucleus. The latter contains contributions of the type

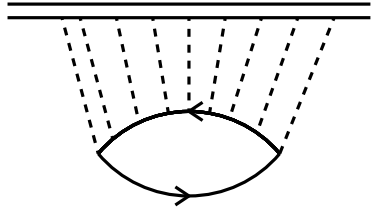

and has singularities connected with atomic states

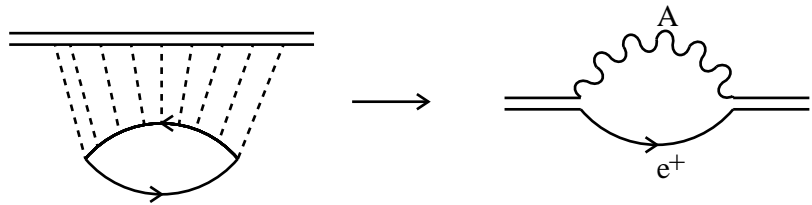

This means, that the Green function of the nucleus will contain this loop, which, definitely, has an imaginary part, because the nucleus becomes heavier than the atom and the positron. It gives us a contribution which describes the instability and results in an exponent $e^{-\Gamma t}$ corresponding to the decay of the nucleus

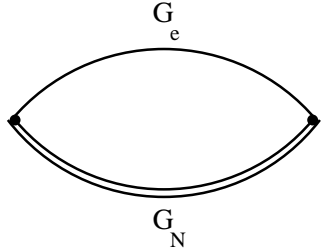

Hence, $e^{\Gamma t}$ will cancel and

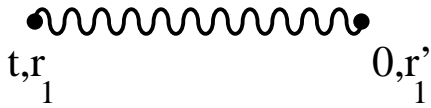

will be proportional to $e^{-i(\omega_0+M)t}$ which corresponds to the propagation of an atomic state with real energy $\omega_0 + M$

So, indeed, we have shown, that in this case the bound state is stable not only due to the existence of binding forces, but also because of the Pauli principle, which reflects the antisymmetry between our electron and the electron in the vacuum. In the electron-nucleus scattering amplitude (4) asymptotically (i.e. at large $t$) only the part

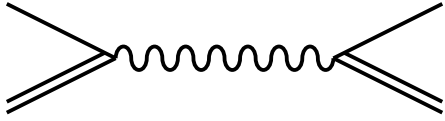

survives in the sense

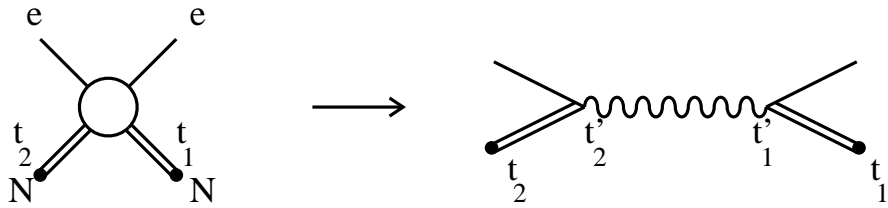

because the Green function of the nucleus decays always exponentially, and it has to be compensated by the increase of the electron Green function. This is possible only, if

$$\Gamma(t_2 - t'_1) < 1 \ , \Gamma(t'_1 - t_1) < 1 \ , t'_2 - t'_1 \to \infty \ .$$

Let us see now, how we have lost the Pauli principle in discussing Feynman Green functions. This can be understood immediately. Our aim is to derive the Feynman prescription from the Dirac picture. In the latter we have negative levels filled up, and we look for the propagation of additional particles.

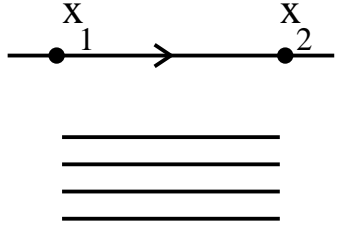

How can we derive Feynman rules from this ? It will be, indeed, very simple.

In this picture our Green function

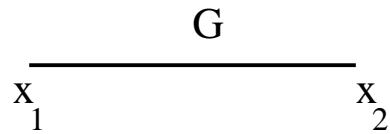

has, of course, to be retarded :

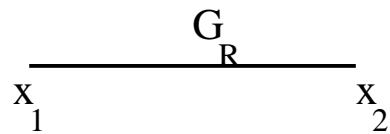

The real propagation is second order in external field :

$$\underset{x_1 \quad A \qquad\quad A \quad x_2}{\times \!\!\!\!\!\!\xrightarrow{\quad G_R \quad}\!\!\!\!\!\! \times} \qquad (20)$$

We start with the usual calculation, but we have to include the Dirac sea. The external field A is acting not only on the particles, but also on the Dirac sea. This means, that we have to consider our particle and a particle from the Dirac see ;

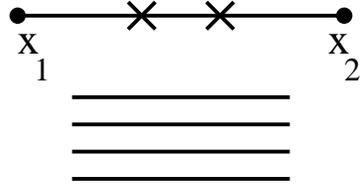

will be a possible diagram. However, we can write also

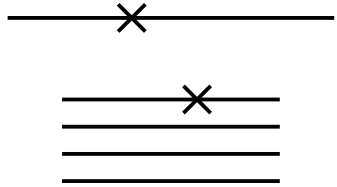

and consider the diagrams

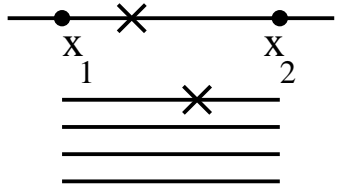

(21)

and

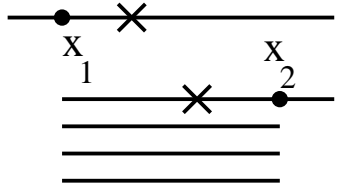

$$\quad (22)$$

Averaging over all particles in the Dirac sea, the diagram (21) leads to

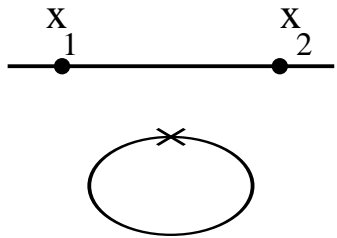

which corresponds to the first order contribution to vacuum polarization and is in fact zero. The diagram (22) gives

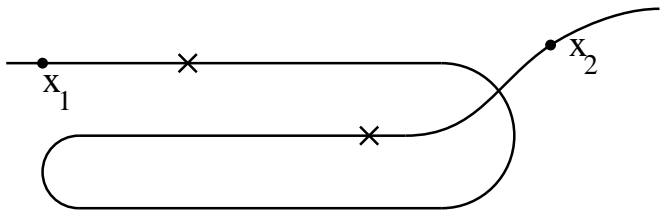

and we get

$$\begin{array}{c}\overset{\times\qquad\times}{\underset{x_1\qquad G_-\qquad x_2}{\rule{3cm}{0.4pt}}}\end{array} \qquad (23)$$

summing all negative energy levels. Adding the diagram with $G_R$ (20) to that with $G_-$ (23), we obtain the Feynman diagram. By these two diagrams, however, symmetry is not reflected. In order to be symmetrical, we have to add which means that the external field is interacting twice with the vacuum. But averaging this, we get i.e. we have to add a diagram which corresponds to vacuum polarization in external field. In other words, the Pauli principle is taken into account by adding vacuum polarization to the normal Feynman propagator.

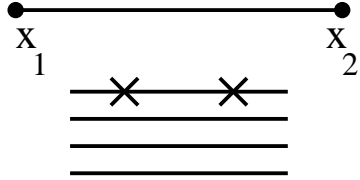

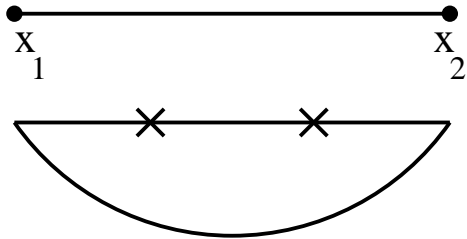

It is, of course, always necessary to consider vacuum polarization, but sometimes it is relevant, sometimes not. In the case of weak coupling the vacuum polarization was real and, when added to the Feynman propagator, did not change the picture essentially. But in our case, in the case of strong coupling, the vacuum polarization becomes complex and because of this, it defines everything.

In conclusion, let us make it clear, which region of distances between the nucleus and the electron is important in our calculation.

As it is shown in [2], the solution $\psi_\omega(r)$ of the Dirac equation (10) is

$$\psi_\omega(r) \sim \frac{1}{\sqrt{r}} cos \left[ \sqrt{(Z\alpha)^2 - 1} \ell n \frac{r}{r_0} + \delta \right] \qquad (24)$$

in the region

$$r_0 << r << \left| \frac{1}{\omega} \right| \quad ,$$

where $r_0$ is the radius of the nucleus. If $|\omega| r_0 << 1$, the function $\psi_\omega(r)$ will oscillate and at $r_0 \to 0$ this corresponds to "falling into the center". If $r_0$ is finite, the "falling into the center" does not occur. However, the existence of oscillations is a sign which indicates that the levels passed through the point $\omega = -m$.

The number of oscillations $n$ in the region $r_0 < r < \frac{1}{m}$ which determines the number of levels passing through is defined by the condition

$$\sqrt{(Z\alpha)^2 - 1} \, \ell n \frac{1}{mr_0} + \delta = n\pi \quad . \qquad (25)$$

At $n=1$ formula (25) provides the condition for the charge to be supercritical (with $r_0$ and $m$ finite) and shows, that the region where the supercritical phenomenon occurs is

$$r_0 << r \leq \frac{1}{m} \ .$$

**Acknowledgements**

The text of these lectures was written by Julia Nyiri. She substantially expanded and edited the original notes she took during the lectures. I am very grateful to Ph. Boucaud, A. Dudas and J. Mourad who formated the text and the figures. Also, I would like to thank the Lab. de Physique Theorique et Hautes Energies, and especially A. Capella, M. Fontannaz, A. Krzywicki, D. Schiff and Tran Than Van for their hospitality, and the warm and inspiring atmosphere during my stay at Orsay.